\documentclass{article}
\usepackage{spconf,amsmath,graphicx}

\usepackage{psfrag}

\usepackage[absolute]{textpos}

\newcommand{\copyrightstatement}{
    \begin{textblock}{15}(0.5,0.3)    % tweak here: {box width}(leftposition, rightposition)
         \noindent
         \centering
         \textblockcolour{white}
         \footnotesize
         \copyright 2024 IEEE. Personal use of this material is permitted. Permission from IEEE must be obtained for all other uses, in any current or future media, including reprinting/republishing this material for advertising or promotional purposes, creating new collective works, for resale or redistribution to servers or lists, or reuse of any copyrighted component of this work in other works
    \end{textblock}
}

\title{Energy Reduction Opportunities in HDR Video Encoding}
\name{Christian Herglotz$^1$, Steven Le Moan$^2$, Alexandre Mercat$^3$}
\address{$^1$Brandenburgische-Technische Universität Cottbus-Senftenberg (BTU),\\ Chair of Computer Engineering, Cottbus, Germany\\
$^2$Norwegian University of Science and Technology (NTNU),
	Colourlabs, 	Gj{\o}vik, Norway\\
	$^3$Tampere University (TAU), 
Ultra Video Group,
Tampere, Finland	}

\begin{document}
%\ninept
%

\copyrightstatement

\maketitle
\begin{abstract}
This paper investigates the energy consumption of video encoding for high dynamic range videos. Specifically, we compare the energy consumption of the compression process using 10-bit input sequences, a tone-mapped 8-bit input sequence at 10-bit internal bit depth, and encoding an 8-bit input sequence using an encoder with an internal bit depth of 8 bit. We find that linear scaling of the luminance and chrominance values leads to degradations of the visual quality, but that significant encoding complexity and thus encoding energy can be saved. An important reason for this is the availability of vector instructions, which are not available for the 10-bit encoder. Furthermore, we find that at sufficiently low target bitrates, the compression efficiency at an internal bit depth of 8 bit exceeds the compression efficiency of regular 10-bit encoding. 
\end{abstract}
\section{Introduction}
\label{sec:intro}

In recent years, the use and production of high dynamic range (HDR) video content increased substantially. One major reason for this development is that state-of-the-art, standard dynamic range (SDR) cameras can be extended with HDR capabilities by simple software updates \cite{Meuel18}, such that there is no necessity to develop and establish costly new cameras. The second major reason is the success of organic light emitting diode (OLED) displays, which have entered the mass market providing natively higher dynamic ranges than legacy liquid crystal displays (LCDs) \cite{Luo15}. Consequently, HDR imaging has become an interesting new feature for end users. 

A downside to this development is the increased processing cost because of a higher bit depth of the source videos. Instead of using 8 bit representations in common YUV formats, 10 bits must be processed. Due to specifications of current processors, as a consequence, the internal calculations performed in the compression stage must be performed on 16 bit word lengths instead of 8 bit, because 10 bit processing is not available by the definition of instruction set architectures (ISAs). Hence, software encoders and decoders for 10 bit HDR videos include a significant, redundant computational overhead of 6 bits in terms of both processing and memory, which can cause a significantly increased processing complexity and energy consumption. 

Performing a literature review, one can find that up to now, the complexity and energy consumption of HDR compression has drawn little attention. For example, for software decoders, the difference in energy consumption between standard dynamic range (SDR) and HDR was investigated in \cite{Kraenzler19} and it was found that on average, HDR decoding consumes $55\%$ more energy than SDR using the same content at a reduced dynamic range.  Fu et al. \cite{Fu18} introduced a fast algorithm for encoding HDR videos using the scalable HEVC framework, achieving significant improvement by incorporating depth information from coding tree units and training classifiers based on luminance levels. Their method resulted in a $43\%$ reduction in encoding time with a minimal $0.54\%$ increase in bit rate. Concerning tone mapping procedures, Ou et al. \cite{Ou21} proposed an energy-efficient tone mapping processor and Xiang et al. \cite{Xiang20} performed an in-depth study of the energy efficiency of different tone mapping implementations on hardware. Le Meur et al. \cite{LeMeur2024} proposed a unified framework for energy-aware encoding and display of HDR content targeting a reduced screen brightness. Concerning the energy consumption of encoding, Chachou et al. compared various software encoders \cite{Chachou2023} and Katsenou et al. \cite{Katsenou2022} investigated the performance of various encoders comparing the rate, distortion, and energy consumption in more detail, but did not consider HDR content specifically. 

In a study by the Shift project \cite{ShiftFull19}, it was found that global online video communications nowadays account for a significant amount of greenhouse gas emissions. Especially for online video services such as social networks, the encoding and transcoding process of videos contribute substantially to the overall energy consumption of online video systems \cite{Herglotz23}. Hence, to reduce the impact of video communications, it is effective and helpful to optimize or omit redundant and energy-intensive processes. In preceding work, we investigated the impact of resolution changes \cite{Herglotz19b} and frame rate changes \cite{Herglotz23a} on the visual quality of videos and the energy consumption of corresponding hardware devices. In this paper, we investigate the impact of the bit depth on the visual quality and the energy consumption for encoding HDR videos.

We study the influence of HDR content on encoding procedures and analyze the impact of compression on bitrate, visual quality, encoding complexity, and encoding energy consumption. We find the following three conclusions when scaling an HDR sequence to SDR and encoding with 8 bit instead of 10 bit: (1) at high quantization parameters (QPs), encoding with an 8 bit tone-mapped representation can lead to an improved rate-distortion performance when inverting the scaling after the decoding process; (2) at a constant visual quality, the encoding energy consumption at 8 bit is more than $75\%$ lower than at 10 bit, which is caused both by single-instruction-multiple-data (SIMD) instructions and smaller word lengths; (3) at 8 bit encoding, as expected, a certain minimum quality degradation is unavoidable.

%Next to In this direction, we study The contributions of this paper are as follows: 
%\begin{itemize}
%\item A comparison of the compression performance using different input bit depths, 
%\item a comparison of the compression performance at different internal bit depths,
%\item a study on the energy consumption  
%\end{itemize}

\begin{figure*}[ht]
\centering
\psfrag{1}[c][c]{10-bit input sequence}
\psfrag{A}[l][l]{\textbf{10-10}}
\psfrag{B}[l][l]{\textbf{8-10}}
\psfrag{C}[l][l]{\textbf{8-8}}
\psfrag{2}[c][c]{10-bit output sequences}
\psfrag{E}[c][c]{Encoder}
\psfrag{a}[c][c]{\small 10 bit}
\psfrag{b}[c][c]{\small 8 bit}
\psfrag{D}[c][c]{Decoder}
\psfrag{r}[c][c]{$>>2$}
\psfrag{l}[c][c]{$<<2$}
\includegraphics[width=.8\textwidth]{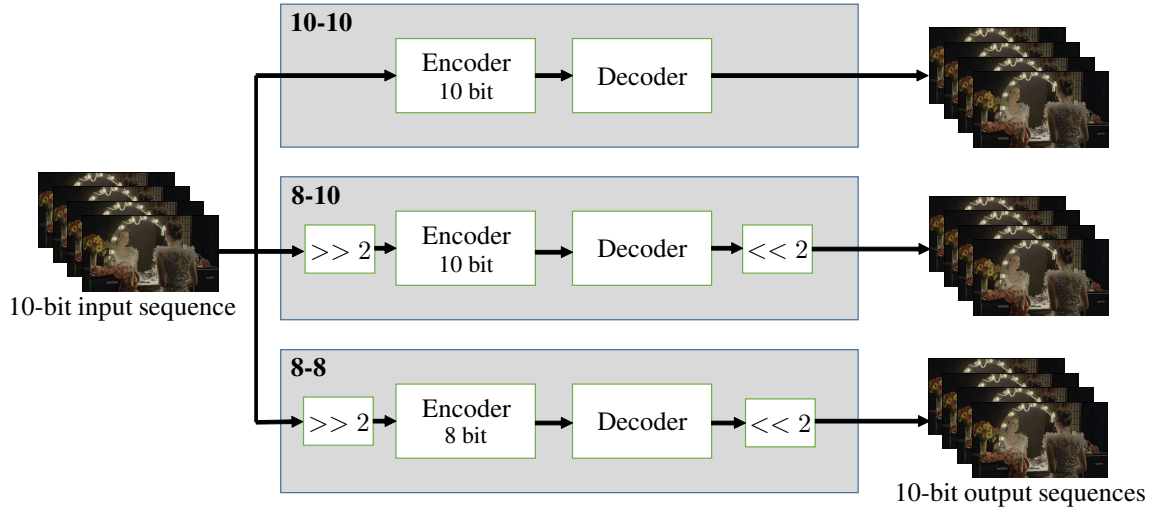}
\vspace{-.4cm}
\caption{Encoding procedures for HDR input videos. For the top and the center variant (10-10, 8-10), an encoder with an internal bit depth of 10 bit is used. The bottom variant uses an 8-bit encoder. For the center and the bottom variant, the 10 bit input video is quantized to an 8 bit input video (indicated by $>>2$). Furthermore, after decoding, the 8-bit output videos are rescaled to 10 bit ($<<2$).  }
\label{fig:encProcess}
\end{figure*}

This paper is organized as follows. First, Section \ref{sec:enc} introduces our encoding configurations and setups. Section~\ref{sec:setup} explains evaluation procedures including quality indices and energy consumption assessment methods. Then, Section \ref{sec:eval} performs an in-depth evaluation of the compression performance and the energy consumption of the different encoder variants. Section~\ref{sec:disc} discusses the results and Section \ref{sec:concl} concludes this paper.

\section{Encoder Setup}
\label{sec:enc}

Fig.~\ref{fig:encProcess} illustrates the three encoding setups to assess the compression performance and energy consumption of HDR encoding. We investigate three variants: First, we perform regular encoding of 10 bit source sequences using encoding at an internal bit depth of 10 bits (10-10), which we can see at the top. This means that during encoding, all intermediate pixel values are stored in a 10 bit format, increasing the accuracy of intermediate calculations. 

 Second, as indicated by the central part in Fig.~\ref{fig:encProcess}, we tone-map the 10 bit source sequences by linear scaling to an 8 bit representation. To this end, we map the input integer data range $[0, 2^{10}-1]$ to a floating point range of $[0,1]$, scale up to the corresponding 8-bit integer range $[0,255]$, and round the resulting pixels values. The corresponding quantization noise power is approximately $1.33$ in terms of mean square error (MSE) in the 10-bit range. Then, we encode the 8-bit representation using an encoder with an internal bit depth of 10 bit (8-10)\footnote{Note that for 8-bit source sequences, this approach is the standard approach in more recent codecs such as Versatile Video Coding (VVC) \cite{ITU_VVC}.}. To restore the original dynamic range with 10 bits, we perform the inverse scaling procedure, which corresponds to appending two zero-valued bits. 
 
Third, as indicated in the bottom part of Fig.~\ref{fig:encProcess}, we perform the same scaling procedures as in the second variant but use a regular 8-bit encoder with an internal bit depth of 8 to encode the scaled source sequence (8-8). At the output of all variants, we obtain 10-bit representations of the sequences that we can compare to the original source sequences using full-reference quality indices.

We use Kvazaar \cite{Viitanen16}, an open source HEVC encoder for our experiments because parts of the codebase are optimized and, unlike reference encoders such as the HM encoder \cite{HM}, targets real-time capable and power-efficient applications. Note that the 10-bit encoder executable is different from the 8-bit encoder executable. The former is compiled with an internal bit depth of 10 bit and the latter with an internal bit depth of 8 bit. As a consequence, pixel values are stored on 8 bits during encoder runtime for the 8-bit encoder, while for the 10-bit encoder, pixel values are stored on 16 bits. Furthermore, the 8-bit encoder makes extensive use of SIMD instructions in various compression tools such as the sample adaptive offset filter (SAO) and the calculation of the sum of squared errors (SSE). 

To extract the complexity impact of SIMD instructions, we perform another study disabling these instructions during the encoding process. Note that the resulting bit streams are exactly the same as in the SIMD-enabled encoder, because the calculations are identical. 

The encoding configuration is the standard configuration at medium preset with quantization parameters (QPs) 12, 17, 22, 27, 32, 37, where we decided to extend the QP range to lower QPs, which retains more information from the extended bit depth of 10 bit. 

The set of input sequences consists of the HDR sequences at high-definition (HD) resolution from the JVET HDR common test conditions \cite{JVET_T2011}. It includes eight sequences provided in the BT.2020 format \cite{Sugawara14}, see Table~\ref{tab:intersections}. The chromatic components are subsampled in the 4:2:0 format. 

\section{Evaluation Procedures}
\label{sec:setup}

We use three types of metrics in this paper: Rate, quality, and complexity. The rate is the average bitrate in bits per second as reported by the encoder. The visual quality is indicated by two different metrics: First, the peak sigal-to-noise ratio (PSNR), which is based on the mean square error (MSE) between the source sequence and the reconstructed sequence. The PSNR is calculated separately for each color component Y, U, and V and averaged to the YUV-PSNR
\begin{equation}
\mathrm{PSNR}_\mathrm{YUV} = \frac{1}{8}\left(6\mathrm{PSNR}_\mathrm{Y} + \mathrm{PSNR}_\mathrm{U} + \mathrm{PSNR}_\mathrm{V}\right), 
\label{eq:PSNR}
\end{equation}
which we report in the evaluation section. 

Second, we report average quality values as calculated by the High Dynamic Range Visible Difference Predictor (HDR-VDP) \cite{Mantiuk23,Mantiuk11}, which estimates the perceived quality of HDR images. The quality as indicated by this metric is defined in the range $[0,10]$ with $10$ corresponding to the highest quality.  HDR-VDP was proven to provide a high correlation with subjective scores. For the calculation of this full-reference metric, we used version v3.0.7 with a viewing distance corresponding to a density of $30$ pixels per degree viewing angle, CPU processing. We assume brightness conversion following the BT.2020 standard \cite{Sugawara14} with an OLED screen as the target display. Note that a perceptual metric for HDR videos (HDR-VDP is trained on images) could not be found in the literature and will be investigated in future work.

Concerning the complexity and the energy consumption of the encoding process, we perform dedicated measurements on two platforms. The first platform is a Windows PC with an Intel(R) Core(TM) i7-8700 CPU @ 3.20GHz. We monitor the complexity by measuring the single-core CPU time, which is highly correlated with the energy consumption \cite{Ramasubbu22a}. 

The second platform is a Linux-based system with an Intel(R) Core(TM) i5-10505 CPU @ 3.20GHz. We perform single-core encoding time measurements in the same way as for the Windows system. In addition, we conduct energy measurements of the single-core encoding using the internal energy measurement circuitry Running Average Power Limit (RAPL) \cite{David10}. We perform each encoding measurement five times to allow reporting of mean values and remove outliers. Note that our energy measurements include both the static energy consumption (idle consumption) as well as the dynamic energy consumption.

For decoding, we use the the open source OpenHEVC decoder \cite{OpenHEVC} included in the FFmpeg framework \cite{FFmpeg}, which automatically chooses the internal bit depth depending on the high-level information encoded in the bit stream.

\section{Evaluation}
\label{sec:eval}

\begin{figure*}[ht]
\centering
\psfrag{015}[tc][c]{ Relative Bitrate Difference (\%)}%
\psfrag{016}[bc][tc]{ PSNR$_\mathrm{YUV}$ in dB}%
\psfrag{017}[tc][tc]{ Bitrate in kbps}%
\psfrag{018}[bc][tc]{ PSNR$_\mathrm{YUV}$ in dB}%
\psfrag{000}[ct][ct]{ $- 100$}%
\psfrag{001}[ct][ct]{ $0$}%
\psfrag{002}[ct][ct]{ $100$}%
\psfrag{003}[ct][ct]{ $200$}%
\psfrag{004}[ct][ct]{ $300$}%
\psfrag{005}[ct][ct]{ $400$}%
\psfrag{006}[ct][ct]{ $500$}%
\psfrag{007}[ct][ct]{ $600$}%
\psfrag{011}[ct][ct]{ $1$}%
\psfrag{008}[rc][rc]{ $40$}%
\psfrag{009}[rc][rc]{ $45$}%
\psfrag{010}[rc][rc]{ $50$}%
\psfrag{012}[rc][rc]{ $40$}%
\psfrag{013}[rc][rc]{ $45$}%
\psfrag{014}[rc][rc]{ $50$}%
\psfrag{Points Aaaaaaaaaaaa}[l][l]{\small 10-10}%
\psfrag{Points B}[l][l]{\small 8-10}%
\psfrag{Points C}[l][l]{\small 8-8}%
\psfrag{RCD AB}[l][l]{\small RCD}%
\psfrag{RCD AC}[l][l]{\small BD-Rate}%
\includegraphics[width=\textwidth]{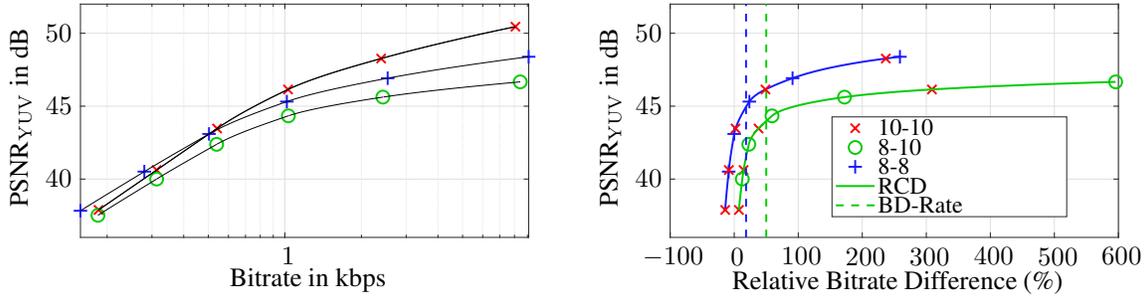}
\vspace{-.4cm}
\caption{Interpolated RD curves (left) and RCDs (right) for the ``Starting'' sequence. The colors indicate different encoder variants from Fig.~\ref{fig:encProcess}. Solid lines correspond to interpolated curves, where the green line on the right is the relative curve difference between variant 10-10 and 8-10; and the blue line between 10-10 and 8-8. The vertical dashed lines indicate the BD-rate values. }
\label{fig:RD1}
\vspace{-.3cm}
\end{figure*}

Our evaluation covers three major aspects: First, we study the rate-distortion performance, second, the encoding energy consumption, and third, the impact of SIMD instructions. 
\subsection{Rate-Distortion Performance}
\label{sec:RDperf}
Fig.~\ref{fig:RD1} illustrates the rate-distortion (RD) curves for the ``Starting'' sequence. 
The graph on the left depicts RD points for different encoder variants: 10-bit to 10-bit (10-10) as red crosses, 8-bit to 10-bit (8-10) as green circles, and 8-bit to 8-bit (8-8) as blue plus signs. Akima interpolation is used to generate smooth RD curves. This interpolation method was selected due to its superior accuracy in matching the actual intermediate RD points compared to other methods \cite{Herglotz24}. The results show that the 10-10 encoder variant achieves the highest PSNR values, exceeding $50\,$dB, which aligns with expectations given its use of the original source bit depth throughout the encoding process. Conversely, reducing the bit depth to 8 bits, in both the 8-10 and 8-8 variants, results in noticeable distortions, with PSNR dropping below $48\,$dB.

The RD performance comparison indicates that the 8-10 variant underperforms relative to the 10-10 variant, as its curve is consistently lower. Interestingly, at PSNR values below 43 dB, the 8-8 variant outperforms the 10-10 variant, suggesting that for lower target qualities, the advantage of a higher internal bit depth diminishes, especially when high quantization parameters lead to the quantization of the least significant bits.

The right graph illustrates the relative curve differences (RCDs) \cite{Herglotz24} between variants. For a fixed PSNR, the RCD illustrates the relative rate difference when switching from one encoder configuration to another. The green curve represents the relative bitrate increase required by the 8-10 variant over the 10-10 variant to achieve the same PSNR, with all values being positive, confirming the superior RD performance of the 10-10 variant. This difference widens at higher PSNR levels. The Bj{\o}ntegaard-Delta (BD) rate, marked by a vertical dashed green line, yields $50\%$.

The blue RCD curve compares the 10-10 variant against the 8-8 variant. Here, the 10-10 variant performs better at high PSNRs (relative bitrate difference is positive). However, below $43\,$dB, the relative bitrate difference turns negative, indicating the 8-8 variant's superior performance in this lower quality range. The overall BD-rate for this comparison is $18.5\%$, with the positive value reflecting the 10-10 variant's efficiency at high PSNR values. The largest rate savings of the 8-8 variant with respect to 10-10 is achieved at the lowest PSNR with a value of $-14\%$. 

\begin{figure*}
\centering
\psfrag{017}[tc][tc]{ Bitrate in kbps}%
\psfrag{015}[tc][c]{ Relative Rate Difference (\%)}%
\psfrag{016}[bc][bc]{ HDR-VDP}%
\psfrag{018}[bc][bc]{ HDR-VDP}%
\psfrag{000}[ct][ct]{ $- 40$}%
\psfrag{001}[ct][ct]{ $- 20$}%
\psfrag{002}[ct][ct]{ $0$}%
\psfrag{003}[ct][ct]{ $20$}%
\psfrag{004}[ct][ct]{ $40$}%
\psfrag{005}[ct][ct]{ $60$}%
\psfrag{010}[ct][ct]{$1$}%
\psfrag{006}[rc][rc]{ $8$}%
\psfrag{007}[rc][rc]{ $8.5$}%
\psfrag{008}[rc][rc]{ $9$}%
\psfrag{009}[rc][rc]{ $9.5$}%
\psfrag{011}[rc][rc]{ $8$}%
\psfrag{012}[rc][rc]{ $8.5$}%
\psfrag{013}[rc][rc]{ $9$}%
\psfrag{014}[rc][rc]{ $9.5$}%
\psfrag{Points Aaaa}[l][l]{10-10}%
\psfrag{Points B}[l][l]{8-10}%
\psfrag{Points C}[l][l]{8-8}%
\psfrag{RCD AB}[l][l]{RCD}%
\psfrag{RCD AC}[l][l]{BD-Rate}%
\includegraphics[width=\textwidth]{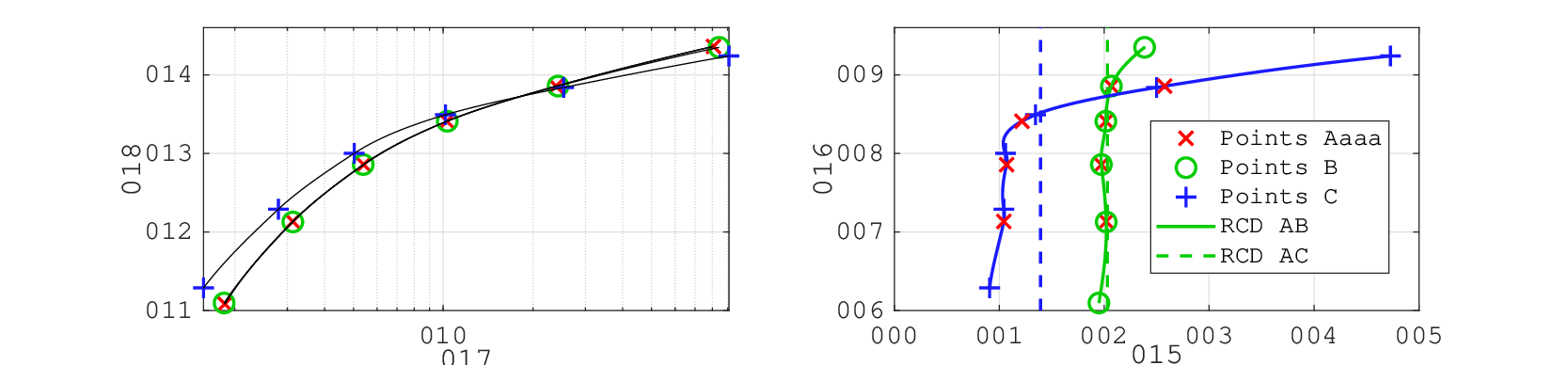}
\vspace{-.5cm}
\caption{Interpolated curves (left) and RCDs (right) for the ``Starting'' sequence. The illustration is the same as in Fig.~\ref{fig:RD1}, but the PSNR on the vertical axes is replaced by the HDR-VDP metric \cite{Mantiuk23}.  }
\label{fig:VDP1}
\vspace{-.5cm}
\end{figure*}

Fig.~\ref{fig:VDP1} shows the corresponding results when exchanging the objective quality PSNR with the perceptually meaningful HDR-VDP index. 
In the left plot, we can see that the difference between 10-10 and 8-10 is reduced significantly. This indicates that quantization of least significant bits has a low impact on the perceived quality. This is also reflected by the green RCD on the right, which is very close to a difference of zero with a mean BD-rate of $0.6\%$, which is negligible. 

\begin{table}[h]
\centering
\vspace{-.3cm}
\caption{Detailed performance comparison of the 10-10 variant with the 8-8 variant for all tested HDR sequences.   The second column indicates the PSNR value of the intersection of the RD curves and the third column the corresponding, next input QP value where the 8-8 variant shows a higher compression than the 10-10 variant. The fourth and fifth columns indicate the BD-encoding time (BDT) value for the Windows (W) and Linux (L) system, and the sixth column the BD-encoding energy (BDEE) for the Linux system. }
\label{tab:intersections}
%\begin{tabular}{r||r|r||r|r}
%\hline 
% & \multicolumn{2}{c||}{Bitrate} & \multicolumn{2}{c}{Complexity}\\
%Sequence & PSNR&  $\lceil$QP$\rceil$ & BDT & BDEE\\\hline
%BalloonFestival & 38.51 & 37 & -69.5$\%$ & -76.7$\%$ \\
%Cosmos1 & 34.82 & 27 & -69.4$\%$ & -77.5$\%$ \\
%Hurdles & 40.96 & 22 & -76.4$\%$ & -83.2$\%$ \\
%Starting & 43.05 & 22 & -76.0$\%$ & -82.3$\%$ \\
%FireEater & 42.11 & 32 & -45.7$\%$ & -60.2$\%$ \\
%Market & 39.59 & 22 & -74.9$\%$ & -80.5$\%$ \\
%ShowGirl & 39.88 & 17 & -72.8$\%$ & -80.1$\%$ \\
%Sunrise & 42.92 & 22 & -72.7$\%$ & -78.8$\%$ \\
%\hline
%Average & - & - & \textbf{-69.7$\%$} & \textbf{-77.4$\%$} \\\hline
\begin{tabular}{r||r|r||r|r|r}
\hline
 & \multicolumn{2}{c||}{Bitrate} & \multicolumn{3}{c}{Complexity}\\
Sequence & PSNR &  $\lceil$QP$\rceil$ & BDT$_\mathrm{W}$ & BDT$_\mathrm{L}$ & BDEE$_\mathrm{L}$ \\\hline
BalloonF. & 38.51 & 37 & -69.5$\%$ & -77.7$\%$ & -76.7$\%$ \\
Cosmos1 & 34.82 & 27 & -69.4$\%$ & -78.8$\%$ & -77.5$\%$ \\
Hurdles & 40.96 & 22 & -76.4$\%$ & -84.4$\%$ & -83.2$\%$ \\
Starting & 43.05 & 22 & -76.0$\%$ & -83.5$\%$ & -82.3$\%$ \\
FireEater & 42.11 & 32 & -45.7$\%$ & -63.3$\%$ & -60.2$\%$ \\
Market & 39.59 & 22 & -74.9$\%$ & -81.9$\%$ & -80.5$\%$ \\
ShowGirl & 39.88 & 17 & -72.8$\%$ & -81.4$\%$ & -80.1$\%$ \\
Sunrise & 42.92 & 22 & -72.7$\%$ & -80.2$\%$ & -78.8$\%$ \\
\hline
Average & & & -69.7$\%$ & -78.9$\%$ & -77.4$\%$ \\\hline
\end{tabular}
\vspace{-.4cm}
\end{table}

In contrast, the compression performance of the 8-8 variant clearly outperforms the compression performance of the 10-10 variant, even at higher quality values corresponding to smaller QPs (the marker at a bitrate of $1\,$kbps). On average, 8-8 outperforms 10-10 with BD-rate of $-12\%$, which is a significant gain. 

Analyzing the other sequences, we found similar results. In general, the 10-10 variant is superior to the 8-10 variant. Concerning 10-10 vs. 8-8, the former is superior to the latter at high qualities, but inferior at low qualities. The intersections of the corresponding RD curves
(using HDR-VDP as the distortion metric) is listed in Table~\ref{tab:intersections}. 
We can see that below output PSNR values between $35\,$dB and $43\,$dB, the 8-8 encoder reaches a higher compression performance. For these intersecting points, we also calculated the corresponding next higher QP $\lceil \mathrm{QP}\rceil$, which indicates the smallest QP where the 8-8 variant outperforms the 10-10 variant in terms of compression performance. We observe a range of $17$ to $27$. As both PSNR and QP values show a very high variability, we conclude that the position of the intersection highly depends on the content of the sequence. This observation motivates future work, where the internal bit depth used for encoding could be chosen depending on the content of the sequence.

\subsection{Complexity \& Energy Consumption}
\begin{figure}
\centering
\psfrag{010}[tc][tc]{ Encoding time in s}%
\psfrag{011}[bc][bc]{ HDR-VDP}%
\psfrag{000}[ct][ct]{ $0$}%
\psfrag{001}[ct][ct]{ $200$}%
\psfrag{002}[ct][ct]{ $400$}%
\psfrag{003}[ct][ct]{ $600$}%
\psfrag{004}[ct][ct]{ $800$}%
\psfrag{005}[rc][rc]{ $8$}%
\psfrag{006}[rc][rc]{ $8.5$}%
\psfrag{007}[rc][rc]{ $9$}%
\psfrag{008}[rc][rc]{ $9.5$}%
\psfrag{009}[rc][rc]{ $10$}%
\psfrag{RD Aaaaaa}[l][l]{10-10}%
\psfrag{RD B}[l][l]{8-10}%
\psfrag{RD C}[l][l]{8-8}%
\includegraphics[width=.48\textwidth]{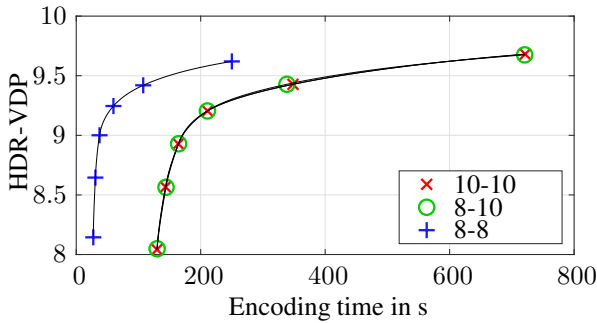}
\vspace{-.5cm}
\caption{Interpolated curves for the ``Starting'' sequence with HDR-VDP as the quality metric and the encoding time on the horizontal axis. For the 8-8 variant, SIMD is enabled.  }
\label{fig:encTimeWin}
\vspace{-.3cm}
\end{figure}

To investigate the complexity of the encoding process, we plot the quality in terms of HDR-VDP versus the encoding time, SIMD on, on the Windows PC in Fig.~\ref{fig:encTimeWin} for the same sequence used before.  
We can clearly see that the 8-8 variant outperforms all 10-bit variants substantially. For all sequences, we calculate BD values by replacing the rate with the encoding time in the BD calculus. Resulting BD-time (BDT) values are listed in the fourth column of Table~\ref{tab:intersections}.We observe that at the same visual quality, the 8-8 variant needs $70\%$ less encoding time on average, within a range between $45\%$ and $77\%$. This indicates that due to the high correlation between encoding time and encoding energy \cite{Ramasubbu22a}, also the encoding energy is reduced significantly. Comparing 10-10 with 8-10, we find that the encoding times only differ marginally. 

On the Linux based system, we verify the results evaluating both the single-core enoding time and the single-core encoding energy. We find that in general, the performance curves are similar to curves for the Windows-based encoder. %as illustrated in Fig.~\ref{fig:statEnergy}. 
%\begin{figure}
%\centering
%\psfrag{009}[tc][tc]{ Encoding Energy in kJ}%
%\psfrag{010}[bc][bc]{ HDR-VDP}%
%\psfrag{000}[tc][tc]{ }%
%\psfrag{001}[ct][ct]{ $0$}%
%\psfrag{002}[ct][ct]{ $5$}%
%\psfrag{003}[ct][ct]{ $10$}%
%\psfrag{004}[ct][ct]{ $15$}%
%\psfrag{005}[ct][ct]{ $20$}%
%\psfrag{006}[rc][rc]{ $8$}%
%\psfrag{007}[rc][rc]{ $9$}%
%\psfrag{008}[rc][rc]{ $10$}%
%\psfrag{RD Aaaa}[l][l]{10-10}%
%\psfrag{RD B}[l][l]{8-10}%
%\psfrag{RD C}[l][l]{8-8}%
%\includegraphics[width=.48\textwidth]{gfx/vdp_energy_v1_matlabfrag}
%\vspace{-.5cm}
%\caption{Interpolated curves for the ``Starting'' sequence with HDR-VDP as the quality metric and the encoding energy measured on the Linux system on the horizontal axis. }
%\label{fig:statEnergy}
%\end{figure}
We calculate corresponding BDT and BD-encoding energy (BDEE) values and list them in columns five and six of Table~\ref{tab:intersections}. Similar to the Windows system, the 8-8 variant is much faster and consumes less encoding energy than the 10-10 variant. Still, the savings are significantly larger ($78.9\%$ time savings for Linux vs. $69.7\%$ savings on Windows), which could be attributed to compiler optimizations or differences in the architecture of  the selected processors. Furthermore, we can see that encoding energy savings are slightly smaller than encoding time savings (by $1.5\%$ on average), which can be explained by the fact that due to parallel processing of data, SIMD operations are more energy-intensive than scalar instructions. 

%The average encoding time and encoding energy savings for all sequences are listed in Table~\ref{tab:intersections}. We can see that on average, the 8-8 encoder is around $70\%$ faster and consumes more than $75\%$ less energy than the 10-10 encoder. 

\begin{figure*}
\centering
\psfrag{019}[tc][tc]{ Encoding energy in kJ}%
\psfrag{017}[tc][c]{ Relative Encoding Energy Difference (\%)}%
\psfrag{018}[bc][bc]{ HDR-VDP}%
\psfrag{020}[bc][bc]{ HDR-VDP}%
\psfrag{000}[ct][ct]{ $- 20$}%
\psfrag{001}[ct][ct]{ $- 10$}%
\psfrag{002}[ct][ct]{ $0$}%
\psfrag{003}[ct][ct]{ $10$}%
\psfrag{004}[ct][ct]{ $20$}%
\psfrag{005}[ct][ct]{ $30$}%
\psfrag{006}[ct][ct]{ $40$}%
\psfrag{011}[ct][ct]{ $1$}%
\psfrag{012}[ct][ct]{ $10$}%
\psfrag{007}[rc][rc]{ $8$}%
\psfrag{008}[rc][rc]{ $8.5$}%
\psfrag{009}[rc][rc]{ $9$}%
\psfrag{010}[rc][rc]{ $9.5$}%
\psfrag{013}[rc][rc]{ $8$}%
\psfrag{014}[rc][rc]{ $8.5$}%
\psfrag{015}[rc][rc]{ $9$}%
\psfrag{016}[rc][rc]{ $9.5$}%
\psfrag{RD Aaaaa}[l][l]{ 10-10}%
\psfrag{RD B}[l][l]{ 8-10}%
\psfrag{RD C}[l][l]{ 8-8 S}%
\psfrag{RD D}[l][l]{ 8-8 D}%
\psfrag{Points A}[l][l]{8-8}%
\psfrag{Points B}[l][l]{8-10}%
\psfrag{RCD AB}[l][l]{8-8 D}%
\psfrag{RCD AC}[l][l]{RCD}%
\psfrag{BD-Rate AB}[l][l]{BDEE}%
\includegraphics[width=\textwidth]{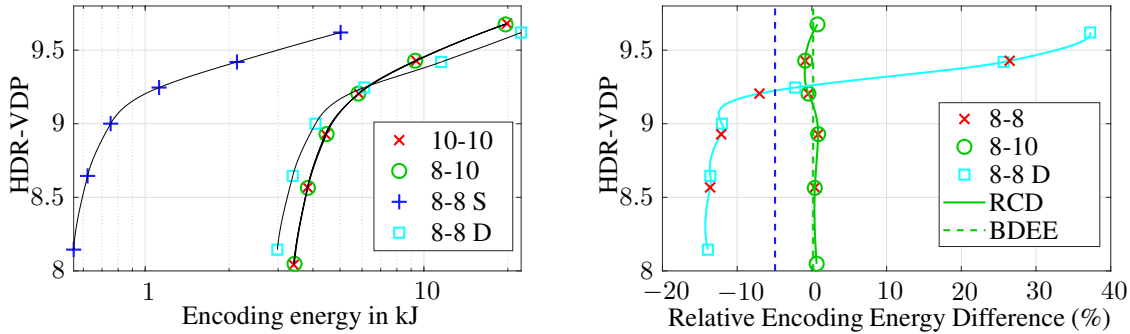}
\vspace{-.5cm}
\caption{Interpolated curves (left) and RCDs (right) for the ``Starting'' sequence, HDR-VDP versus the encoding energy. The 8-8 S variant is the 8-8 encoder with SIMD enabled, 8-8 D with SIMD disabled. .   }
\label{fig:SIMD}
\vspace{-.5cm}
\end{figure*}

An outlier to these observations is the FireEater sequence, which shows lowest savings among all sequences. The reason is that the content of this sequence shows some striking characteristics with extremely bright areas (fire) and extremely dark areas (black night sky), which leads to considerable quality losses as per the HDR-VDP metric when scaling the source to 8 bit. 

\subsection{Impact of SIMD Instructions}

As mentioned before, SIMD instructions lead to significant processing time savings, rendering the complexity comparison between 8-8 (which heavily supports SIMD) and 10-10 (which does not include any SIMD instructions) unfair, because the complexity difference can be caused by manually performed optimizations. Therefore, in another study, we disable SIMD instructions for the 8-8 encoder such that it can be compared to 10-10 using the same level of optimizations. Note that the functionality of the 8-8 encoder with SIMD enabled is identical to the functionality with SIMD disabled, such that the RD performance (cp. Subsection \ref{sec:RDperf}) is the same for both 8-8 variants. 

Therefore, we present the performance in terms of the quality (HDR-VDP) and encoding energy in Fig.~\ref{fig:SIMD}. 
The left plot shows the interpolated curves for the 10-10 variant (red `x'), 8-10 (green `o'), 8-8 with SIMD enabled (blue `+'), and 8-8 with SIMD disabled (cyan squares). We can see that a large part of the energy savings by the 8-8-SIMD variant are lost when disabling SIMD. In fact, with SIMD disabled, the curve shows a similar behavior as RD curves in Fig.~\ref{fig:VDP1}, where 8-8 is superior to 10-10 at low qualities and inferior at high qualities. 

The right part of Fig.~\ref{fig:SIMD} shows corresponding RCD curves indicating that below an HDR-VDP of 9.3, the 8-8 encoder is more efficient than the 10-10 encoder. Across the full range, the BDEE value is $-5\%$. Hence, also in terms of the encoding energy consumption, we conclude that at low target qualities, it is beneficial to encode HDR sequences in a tone-mapped 8-bit representation. 

Although the results indicate that when considering the encoding energy, the 8-8 variant should always be chosen, it should be noted that the 10-10 variant appears to consistently produce a marginally higher visual quality. The significance of this difference in perceived quality should be evaluated via subjective testing in future work. Furthermore, it should be noted that our 10-bit to 8-bit conversion is akin to a very rough tone mapping and further investigations should explore the use of energy-efficient and perception-aware tone mapping methods.

\section{Discussion}
\label{sec:disc}
The results presented in this paper indicate that using an 8-bit encoder instead of a 10-bit encoder can lead to substantial energy savings at little to no quality losses. Furthermore, significant rate savings can be obtained at sufficiently low visual qualities. A major reason it that the 8 bit encoder (variant 8-8) makes substantial use of SIMD optimizations, which are not available for the 10-bit encoder, which explains most of the energy and time savings. However, this does not explain rate savings observed at low visual qualities. Reasons for the superior RD performance in these regions could be sub-optimal RD decisions due to an imperfect relation between the Lagrange multiplier and the input QP \cite{Sullivan98} or differences in the choice of coding modes and the performance of loop filters. Note that we would expect the 10-10 variant to outperform the 8-8 variant for 10-bit input sequences independent from the target bitrate or the target quality. Therefore, analyzing and improving the encoding algorithm of the 10-10 variant correspondingly is an interesting topic for future research.

\section{Conclusions}
\label{sec:concl}
In this paper, we investigated the compression performance and the energy consumption when encoding HDR sequences with different input bit depths. We have found that 8-bit encoding of 10-bit source sequences can lead to bitrate savings and energy savings at low qualities. Additional encoding energy savings of more than $75\%$ can be reached when making use of SIMD optimizations. 

In future work, we will investigate improved encoding decisions for 10-bit encoders targeting to mitigate the superior compression performance of the 8-bit encoder at low qualities. Furthermore, we will combine the approach of bitrate scaling with scaling of other dimensions such as the resolution and the framerate to obtain energy-optimal encoding configurations at the same visual quality. 

%\vfill%\pagebreak

\bibliographystyle{IEEEbib}
\bibliography{lit.bib}

\end{document}